\documentclass[conference]{IEEEtran}
\IEEEoverridecommandlockouts
\usepackage{cite}
\usepackage{amsmath,amssymb,amsfonts}
\usepackage{algorithm}
\usepackage{algpseudocode}
\usepackage{graphicx}
\usepackage{textcomp}
\usepackage{xcolor}
\algrenewcommand\algorithmicrequire{\textbf{Input:}}
\algrenewcommand\algorithmicensure{\textbf{Output:}}
\def\BibTeX{{\rm B\kern-.05em{\sc i\kern-.025em b}\kern-.08em
    T\kern-.1667em\lower.7ex\hbox{E}\kern-.125emX}}

\makeatletter
\newcommand{\linebreakand}{%
  \end{@IEEEauthorhalign}
  \hfill\mbox{}\par
  \mbox{}\hfill\begin{@IEEEauthorhalign}
}
\makeatother

\begin{document}
\title{A Review on Optimality Investigation Strategies for the Balanced Assignment Problem}

\author{\IEEEauthorblockN{1\textsuperscript{st} Anurag Dutta}
\IEEEauthorblockA{\textit{Department of Computer Science and Engineering} \\
\textit{Government College of Engineering \& Textile Technology}\\
Serampore, Calcutta, India \\
anuragdutta.research@gmail.com}
\and 
\IEEEauthorblockN{2\textsuperscript{nd} K. Lakshmanan}
\IEEEauthorblockA{\textit{Department of Mathematics} \\
\textit{Kuwait American School of Education, Salmiya}\\
Hawalli, Kuwait \\
coprime65@gmail.com}
\linebreakand 
\IEEEauthorblockN{3\textsuperscript{rd} A. Ramamoorthy}
\IEEEauthorblockA{\textit{Department of Mathematics} \\
\textit{Velammal Engineering College, Anna University}\\
Chennai, Tamil Nadu, India \\
ramzenithmaths@gmail.com}
\and
\IEEEauthorblockN{4\textsuperscript{th} Liton Chandra Voumik}
\IEEEauthorblockA{\textit{Department of Economics} \\
\textit{Noakhali Science and Technology University}\\
Noakhali, Bangladesh \\
litonvoumik@gmail.com}
\linebreakand 
\IEEEauthorblockN{5\textsuperscript{th} John Harshith}
\IEEEauthorblockA{\textit{Department of Computer Science} \\
\textit{Vellore Institute of Technology University}\\
Vellore, Tamil Nadu, India \\
johnharshith@icloud.com}
\and 
\IEEEauthorblockN{6\textsuperscript{th} John Pravin Motha}
\IEEEauthorblockA{\textit{School of Management Studies}\\
\textit{REVA University, Yelahanka, Bengaluru}\\
Karnataka, India\\
johnpravin.motha@reva.edu.in}
}

\maketitle
% typeset the header of the contribution

\begin{abstract}
Mathematical Selection is a method in which we select a particular choice from a set of such. It have always been an interesting field of study for mathematicians. Accordingly, Combinatorial Optimization is a sub field of this domain of Mathematical Selection, where we generally, deal with problems subjecting to Operation Research, Artificial Intelligence and many more promising domains. In a broader sense, an optimization problem entails maximising or minimising a real function by systematically selecting input values from within an allowed set and computing the function's value. A broad region of applied mathematics is the generalisation of metaheuristic theory and methods to other formulations. More broadly, optimization entails determining the finest virtues of some fitness function, offered a fixed space, which may include a variety of distinct types of decision variables and contexts. In this work, we will be working on the famous Balanced Assignment Problem, and will propose a comparative analysis on the Complexity Metrics of Computational Time for different Notions of solving the Balanced Assignment Problem.
\end{abstract}

\begin{IEEEkeywords}
Combinatorial Optimization, Branch and Bound, Brute Force, Hungarian Algorithm, Artificial Intelligence
\end{IEEEkeywords}

\section{Introduction}
A foundational combinatorial problem formulation is the assignment problem. The challenge, in its simplest and general form, is as described in the following - "There are several agents and tasks in the problem under consideration. Any operative can indeed be delegated to undertake any task, at a cost that varies based on the operative assessment. It is necessary to complete as many tasks as feasible by allocating no more than one operative to each job at hand and no more than one job to each operative, to ensure that the total cost of the assessment is minimised.". Mathematically, the Problem Statement is as, \\\\
\textbf{General Assignment Problem: }\textit{Optimal assignment for $\mathcal{K}$ workers, $\mathcal{W}_i\forall i=1,2,3,...,\mathcal{K}$ having an equal number of Jobs, $\mathcal{J}_i\forall i=1,2,3,...,\mathcal{K}$ with associated Job Cost, $\mathcal{C}_i\forall i=1,2,3,...,\mathcal{K}$\\\\}
A good example for the demonstration of the Problem Statement would follow as, Assume a cab company does have 3 cabs (operatives) obtainable as well as 3 clients (jobs) who want to be scooped up as quickly as possible. The company takes dignity in quick pickups, so the expense of pulling up a specific client for each cab will be determined by the time it necessitates the cab to arrive at the destination port. This is a concern of balanced assignment. Its workaround is that whichever taxi-customer pairing outcomes in the lowest overall cost. \\\\
Numerous business organizations worldwide of trade strive to make the best use of their constrained resources across numerous activities. They can only do so by employing the assignment problem procedure. It is a subset of transportation problems in the trade industry, with the main objective of assigning an equal proportion of beginnings to an equivalent assortment of destinations. This method entails assigning people to diverse programs, job prospects to machineries, and educators to classrooms, among other things. Assessment issues really aren't restricted to commercial enterprises. They can be utilized in a plethora of ways to assist the user in assigning specific individuals to various tasks or responsibilities. The following are some of its primary application areas
\begin{enumerate}
\item It is employed in the production plant to allocate machinery and equipment to different arrangements.
\item It can be employed in advertising firms within which management teams allocate distinct salespeople to different regions.
\item It can be employed to potential competitive advantage to distinct tenderers.
\item It can also be employed in educational domains like, allocation of educators to separate classes.
\item Distinct management consultants are assigned to distinct banking customers in a specific area.
\end{enumerate}
This is a classical problem of Combinatorial Optimization, as we are going to search for the optimal cost value, or in just terms, we will have to search for the Permutation, that will grant us a minimal cost, in total. Now, there are a variety of methods to solve this problem, like the Brute Force Solution, Solution subjecting Hungarian Method, Branch and Bound, etc. We will be discussing about each method in details and will try to pose a comparative study on them based on the notion of Computational Time. 
\section{Brute Force Approach}
The Brute Force solution to this problem is quite understandable from the Problem Statement itself. We will run a selection on the $\mathcal{K}!$ permutations of the arrangement, of assigning a Job, to a Worker, followed by a comparator to generate the minimum of all. So, in total, the number of iterations that, the Approach is supposed to suffer is of the order $O(\mathcal{K}!)$. It is quite a naïve approach, and is perhaps going to take the most time to be executed. Algorithm \ref{alg1} shows the Pseudo Code to the naïve approach.
\begin{algorithm}
\caption{Pseudo Code for the Brute Force Approach}\label{alg1}
\begin{algorithmic}[1]
\Require $\mathcal{M}_{\mathcal{K}\times \mathcal{K}}\ni\mathcal{M}_{\left(i,\ j\right)}= $ $i$'th worker's charge for $j$'th job.
\Ensure $\operatorname*{argmin} {\left(\sum_{i=0}^{\mathcal{K}-1}\left(\sum_{j=0}^{\mathcal{K}-1}\left({\hat{\mathfrak{C}}}_{\left(i,\ \ \ j\right)}\right)\right)\right)}$
      \Function{NaïveApproach}{$\mathcal{M}_{\mathcal{K}\times \mathcal{K}}$}
      \State $\mathcal{L}\left(\mathcal{K}\right)\gets\left\{1,\ 2,\ 3,\ ...,\mathcal{K}\right\}$
      \State $\mathfrak{P}\left(\mathcal{K}!\right)\gets\mathcal{P}\left\{\mathcal{L}\left(\mathcal{K}\right)\right\}$ \Comment{$\mathfrak{P}\left(\mathcal{K}!\right)$ is the List of Permutations of $\mathcal{L}\left(\mathcal{K}\right)$}
      \State $m \gets 0$ \Comment{$m$ stores the Minimal Cost}
      \For{$i=0$ to $\mathcal{K}!$}
	  \State $m \gets $ $min$($m$, $\hat{\mathfrak{C}}$($\mathfrak{P}_i$)) \Comment{$\hat{\mathfrak{C}}\left(.\right)$ is the Cost Function}
      \EndFor
      \EndFunction
\end{algorithmic}
\end{algorithm}
\section{Kuhn – Munkres Algorithm}
It is a polynomial algorithm[1], that guarantees the optimality for the Total Optimal Cost. Firstly, we will have to generate the Cost Matrix, $\mathcal{M}_{\mathcal{K}\times \mathcal{K}}\ni\mathcal{M}_{\left(i,\ j\right)}= $ $i$'th worker's charge for $j$'th job. From the cost matrix, we will have to select $\mathcal{K}$ elements, one from each row, and return the sum of these $\mathcal{K}$ elements. The Algorithm is as follows
\begin{enumerate}
\item Let's for instance, consider, the Initial Cost Matrix 
\begin{equation*}
\mathcal{M}_{\mathcal{K}\times\mathcal{K}}=\left[\begin{matrix}{\hat{\mathfrak{C}}}_{\left(0,\ \ \ 0\right)}&\cdots&{\hat{\mathfrak{C}}}_{\left(0,\ \ \ \mathcal{K}-1\right)}\\\vdots&\ddots&\vdots\\{\hat{\mathfrak{C}}}_{\left(\mathcal{K}-1,\ \ \ 0\right)}&\cdots&{\hat{\mathfrak{C}}}_{\left(\mathcal{K}-1,\ \ \ \mathcal{K}-1\right)}\\\end{matrix}\right] 
\end{equation*}
such that $\forall \  {\hat{\mathfrak{C}}}_{\left(i,\ \ \ j\right)}\geq0$
\item Under a Row Reduction, followed my a Column Reduction we are guaranteed to get atleast one row with atleast one entry as 0. In Row Reduction, we will be subtracting $min({\hat{\mathfrak{C}}}_{\left(i,\ \ \ j\right)})\ \forall\ j\ >0\ \&\ j\ <\mathcal{K}-1$ from the $i$'th row of $\mathcal{M}_{\mathcal{K}\times\mathcal{K}}$, whereas in Column Reduction, we will be subtracting $min({\hat{\mathfrak{C}}}_{\left(i,\ \ \ j\right)})\ \forall\ i\ >0\ \&\ i\ <\mathcal{K}-1$ from the $j$'th column of $\mathcal{M}_{\mathcal{K}\times\mathcal{K}}$.
\item Now, mark all the rows and columns having zeroes with lines, subjected to minimality.
\item If number of lines = $\mathcal{K}$, we have hit the core. Else, we will have to determine the smallest entry not covered by any line. Subtract this entry from each uncovered row, and then add it to each covered column, and return back to Step 3. 
\end{enumerate}
The Algorithm is subjected to be done in a maximum of $\mathcal{K}^3$ iterations, and is of the order, $O(\mathcal{K}^3)$.
\section{Branch and Bound}
A quite effective, method to solve Hard Problems in CS is Branch and Bound[2]. The First thing we have to do to follow this \textit{modus operandi} is to generate the State Space Tree, that represents all possible states, a system can reach out to. The State Space of a discrete system defined by a function $\phi(x)$ can be modelled as a Directed Graph in Dynamical System Theory, where every other potential state of the dynamical is depicted by a node with a directed edge from $a$ to $b$ iff $\phi(a) = b$. After developing the State Space tree, with a fixed branching factor, we reach out to Depth First Search Approach or Breadth First Search Approach, which is used to search for the optimality. If we make use of DFS, such B \& B technique is termed as \textit{LIFO Branch and Bound}, the either being named as \textit{FIFO Branch and Bound}. Well!, we would also make use of \textit{Lowest Cost Branch and Bound}, which indeed is a Heuristic Technique to do the same. Algorithm \ref{alg2} is the Generalized Pseudo Code for B \& B Approach.
\begin{algorithm}
\caption{Pseudo Code for the B \& B}\label{alg2}
\begin{algorithmic}[1]
\Require $\mathcal{M}_{\mathcal{K}\times \mathcal{K}}\ni\mathcal{M}_{\left(i,\ j\right)}= $ $i$'th worker's charge for $j$'th job.
\Ensure $\operatorname*{argmin} {\left(\sum_{i=0}^{\mathcal{K}-1}\left(\sum_{j=0}^{\mathcal{K}-1}\left({\hat{\mathfrak{C}}}_{\left(i,\ \ \ j\right)}\right)\right)\right)}$
      \Function{BranchAndBoundApproach}{$\mathcal{M}_{\mathcal{K}\times \mathcal{K}}$}
      \State $\mathfrak{S}\left(\sum_{i=0}^{\mathcal{K}-1}\mathcal{K}^i\right) \gets $ State Space Tree
      \State $m \gets 0$ \Comment{$m$ stores the Minimal Cost}
      \For{$\delta=0$ to $\mathcal{K}$} \Comment{$\delta$ is the traversal milieu}
	  \State $m \gets $ $min$($m$, $\hat{\mathfrak{C}}\left(\mathfrak{S}_\delta\right)$)) 
      \EndFor
      \EndFunction
\end{algorithmic}
\end{algorithm}
\subsection{FIFO \& LIFO Branch and Bound}
For this Branch and Bound Technique, we make use of the Depth First Search[3] or Breadth First Search[4]. Like, the State Space Tree will be subjected to a DFS. Generation of the State Space Tree is expected to take $O^*(\mathcal{K}^\mathcal{K})$ amortized time. Computational Complexity of DFS or BFS is of the order $O(V + E)$, where $V$ is the number of vertices subjected to the Graph $G$, on which the search is to be applied, and $E$ being the number of edges. 
\textbf{\\Theorem 1. } \textit{For a Finite $\lambda$ - ary tree, $n(E) = n(V) - 1$.\\}
\textbf{\\Proof \\} Let us consider a tree with $n$ vertices.
\begin{figure}[htbp]
\centerline{\includegraphics[width = \linewidth]{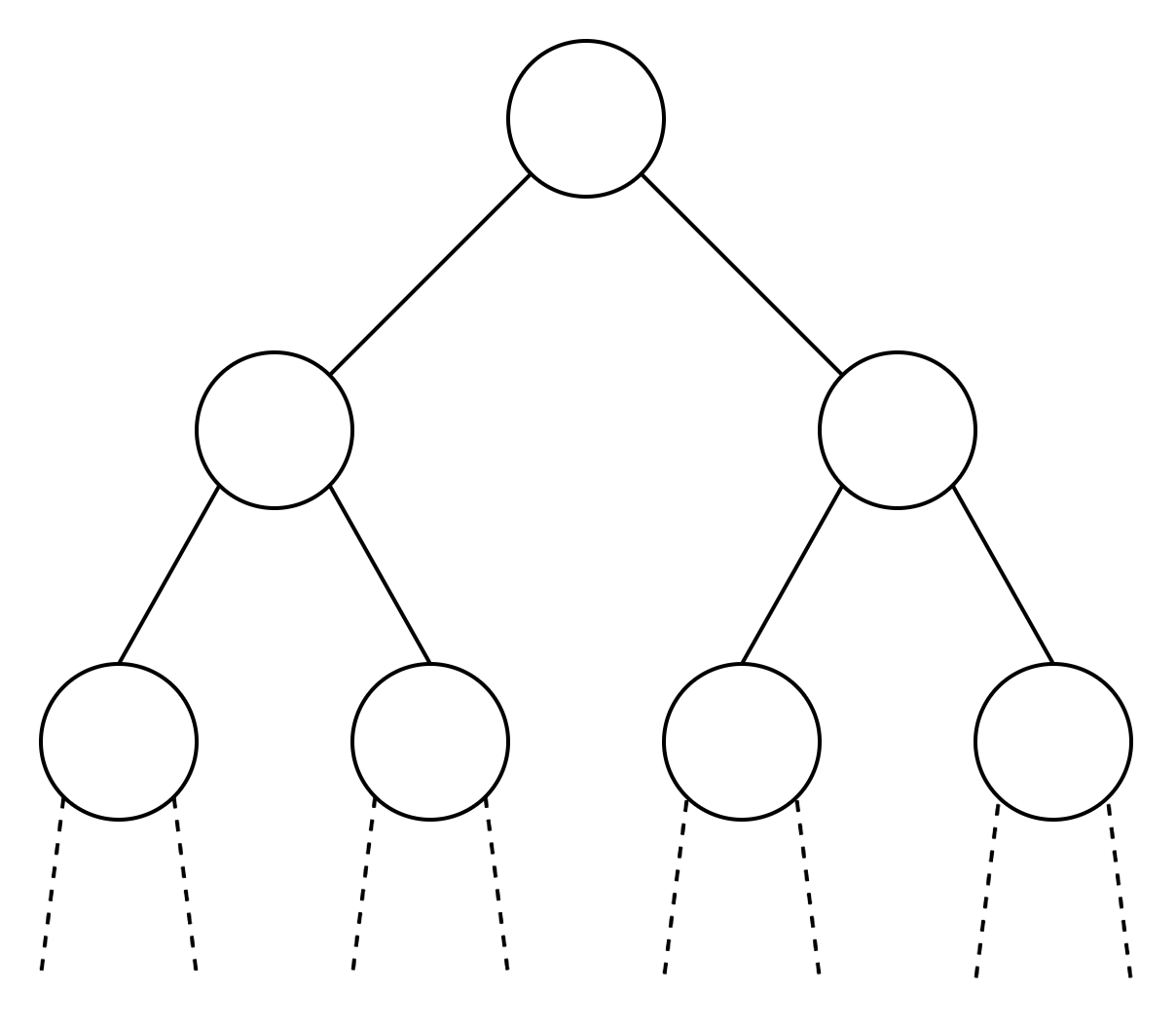}}
\label{fig1}
\caption{A Tree, $T$ with $n$ vertices}
\end{figure} 
\\Let us assume the proposition, $P(n) = n(E) = n-1$ $\forall n \in\mathbb{N}$\\\\
For, $P(n = 1)$, the number of nodes, or vertices is 1, and hence, there is no edges associated with it. Thus, the proposition, $P(n = 1)$ holds true. \\\\
For, $P(n = 2)$, the number of nodes, or vertices is 2, and hence, there is at max, only 1 edge associated with it. Thus, the proposition, $P(n = 2)$ holds true. \\\\
Assume, the proposition $P(n = n)$ is true.  \\\\
Therefore, by Principle of Mathematical Induction, for the proposition, $P(n = n + 1)$, the number of edges will be $n - 1$ + Edge Contribution for adding the new node.\\
Now, for adding a new node, only one edge will be added. Hence, $P(n + 1) = n$\\\\

Hence, we could withstand the proposition $P(n)$. \\\\Straitening on that, we could say that the Computational Complexity of FIFO \& LIFO Branch and Bound would turn out to be 
\begin{equation}
O^\ast(\mathcal{K}!)+O\left(\sum_{i=1}^{\mathcal{K}}\mathcal{K}^i+\sum_{i=0}^{\mathcal{K}}\mathcal{K}^i\right)
\end{equation}
\subsection{Dijkstra's Algorithm}
\textit{Dijkstra's Algorithm}[5] is itself a revolutionary algorithm in the strata of Finding Shortest Path between nodes in a Graph. If we subject that to Cost Function, $\hat{\mathfrak{C}}\left(.\right)$, we are ought to get some good.  Computational Complexity of Dijkstra' is of the order $O(E + Vlog(V))$, where $V$ is the number of vertices subjected to the Graph $G$, on which the search is to be applied, and $E$ being the number of edges. Straitening on Theorem 1, we could claim that the Computational Complexity of B \& B using Dijkstra's Algorithm would turn out to be
\begin{equation}
O^\ast(\mathcal{K}!)+O\left(\sum_{i=1}^{\mathcal{K}}\mathcal{K}^i+\sum_{i=0}^{\mathcal{K}}\mathcal{K}^ilog{\left(\sum_{i=0}^{\mathcal{K}}\mathcal{K}^i\right)}\right)
\end{equation}\\
\subsection{Heuristic Search}
Heuristics are techniques used in optimization techniques and information science to solve problems more efficiently when conventional means are too slow to discover an approximate value or when conventional means are unable to identify any optimal solutions. This is accomplished by exchanging pace for optima, comprehensiveness, accurateness, or finesse. Heuristics, or Informed Search function, is an algorithmic instance that guarantees the best, or the most promising path. The Importance of this Heuristic Function, is to keep a track, of how close, is the current state closer to the Goal State. Instantiating upon that, we could definitely use this Search to search upon the State Space Tree. Computational Complexity of A - Star[6] is of the order $\beta^\Delta$, with $\beta$ being the branching factor[7], and $\Delta$ being the Depth. Straitening on Theorem 1, we could claim that the Computational Complexity[8] of B \& B[9] using A$^*$ Algorithm would turn out to be
\begin{equation}
O^\ast(\mathcal{K}!)+O\left(\mathcal{K}^{\left\lfloor{log}_k{\left(\sum_{i=0}^{\mathcal{K}}\mathcal{K}^i\right)}\right\rfloor}\right)
\end{equation}
\section{Illustrative Example}
In this example, we will try to pose a comparison between the different strategies [10] that may be used for solving the B.A.P. Let's consider a set of 3 Jobs, $\left\{\mathcal{J}_1,\mathcal{J}_2,\mathcal{J}_3\right\}$, and a set of 3 workers, $\left\{\mathcal{W}_1,\mathcal{W}_2,\mathcal{W}_3\right\}$, and suppose their cost of the doing the jobs are as per the cost matrix [11] $\mathcal{M}$. 
\begin{equation*}
\mathcal{M}_{3\times3}=\left[\begin{matrix}9&8&7\\6&5&4\\3&2&1\\\end{matrix}\right]
\end{equation*}
\subsection{Naive Approach}
Let's first try to solve the problem through Naive Approach [12]. There will be 3! ways to solve the problem. The solutions being, 
\begin{enumerate}
\item $\mathcal{J}_1$ being done by $\mathcal{W}_1$; $\mathcal{J}_2$ being done by $\mathcal{W}_2$; and $\mathcal{J}_3$ being done by $\mathcal{W}_3$. Here, Total Cost = $9 + 5 + 1$ = 15
\item $\mathcal{J}_1$ being done by $\mathcal{W}_2$; $\mathcal{J}_2$ being done by $\mathcal{W}_3$; and $\mathcal{J}_3$ being done by $\mathcal{W}_1$. Here, Total Cost = $8 + 4 + 3$ = 15
\item $\mathcal{J}_1$ being done by $\mathcal{W}_3$; $\mathcal{J}_2$ being done by $\mathcal{W}_1$; and $\mathcal{J}_3$ being done by $\mathcal{W}_2$. Here, Total Cost = $7 + 6 + 2$ = 15
\item $\mathcal{J}_1$ being done by $\mathcal{W}_1$; $\mathcal{J}_2$ being done by $\mathcal{W}_3$; and $\mathcal{J}_3$ being done by $\mathcal{W}_2$. Here, Total Cost = $9 + 4 + 2$ = 15
\item $\mathcal{J}_1$ being done by $\mathcal{W}_2$; $\mathcal{J}_2$ being done by $\mathcal{W}_1$; and $\mathcal{J}_3$ being done by $\mathcal{W}_3$. Here, Total Cost = $8 + 6 + 1$ = 15
\item $\mathcal{J}_1$ being done by $\mathcal{W}_3$; $\mathcal{J}_2$ being done by $\mathcal{W}_2$; and $\mathcal{J}_3$ being done by $\mathcal{W}_1$. Here, Total Cost = $7 + 5 + 3$ = 15
\end{enumerate}
Therefore, considering all the possible solutions,
\begin{equation*}
argmin{\left(\sum_{i=0}^{2}\left(\sum_{j=0}^{2}\left({\hat{\mathfrak{C}}}_{\left(i,\ \ \ j\right)}\right)\right)\right)}=15
\end{equation*}
\subsection{Kuhn – Munkres Algorithm}
Now, we would attempt to solve the illustration using the Hungarian Approach. 
\begin{enumerate}
\item The first step would be to search out the row minima(s), the respective row minima(s) would be 7, 4, and 1 respectively, correspondence to
\begin{equation*}
\mathcal{M}_{3\times3}=\left[\begin{matrix}9&8&\fbox{$7$}\\6&5&\fbox{$4$}\\3&2&\fbox{$1$}\\\end{matrix}\right]
\end{equation*}
Now, we would subtract the respective values of the row minors [13] from the entries in the rows. The modified Cost Matrix, becoming, 
\begin{equation*}
\mathcal{M}_{3\times3}=\left[\begin{matrix}2&1&0\\2&1&0\\2&1&0\\\end{matrix}\right]
\end{equation*}
\item The following step would be to subtract the respective values of column minors from the entries in the column, the respective column minors being, 2, 1, and 0 respectively, correspondence to
\begin{equation*}
\mathcal{M}_{3\times3}=\left[\begin{matrix}2&1&0\\2&1&0\\\fbox{$2$}&\fbox{$1$}&\fbox{$0$}\\\end{matrix}\right]
\end{equation*}
Following the aforementioned reduction, our Cost Matrix turns out to be, 
\begin{equation*}
\mathcal{M}_{3\times3}=\left[\begin{matrix}0&0&0\\0&0&0\\0&0&0\\\end{matrix}\right]
\end{equation*}
\item And, finally, we need to cover all zeros with a minimum number of lines. We could do so as, $\left[\begin{matrix}\fbox{$0$}&0&0\\0&\fbox{$0$}&0\\0&0&\fbox{$0$}\\\end{matrix}\right]$, or, $\left[\begin{matrix}0&0&\fbox{$0$}\\0&\fbox{$0$}&0\\\fbox{$0$}&0&0\\\end{matrix}\right]$, or, $\left[\begin{matrix}\fbox{$0$}&\fbox{$0$}&\fbox{$0$}\\0&0&0\\0&0&0\\\end{matrix}\right]$, or, $\left[\begin{matrix}0&0&0\\\fbox{$0$}&\fbox{$0$}&\fbox{$0$}\\0&0&0\\\end{matrix}\right]$, or, $\left[\begin{matrix}0&0&0\\0&0&0\\\fbox{$0$}&\fbox{$0$}&\fbox{$0$}\\\end{matrix}\right]$, or, $\left[\begin{matrix}\fbox{$0$}&0&0\\\fbox{$0$}&0&0\\\fbox{$0$}&0&0\\\end{matrix}\right]$, or, $\left[\begin{matrix}0&\fbox{$0$}&0\\0&\fbox{$0$}&0\\0&\fbox{$0$}&0\\\end{matrix}\right]$, or, $\left[\begin{matrix}0&0&\fbox{$0$}\\0&0&\fbox{$0$}\\0&0&\fbox{$0$}\\\end{matrix}\right]$. In each of these arrangements, Total Cost = 15. 
\end{enumerate}
Therefore, considering all the possible solutions,
\begin{equation*}
argmin{\left(\sum_{i=0}^{2}\left(\sum_{j=0}^{2}\left({\hat{\mathfrak{C}}}_{\left(i,\ \ \ j\right)}\right)\right)\right)}=15
\end{equation*}
\subsection{Branch and Bound}
The same cost matrix can be minimalized using State Space Tree uncoiling [14]. In such a scenario [15], Figure 2, shows a Similar Space Tree image [16]. 
\begin{figure*}[ht]
\centering
\centerline{\includegraphics[width = \linewidth]{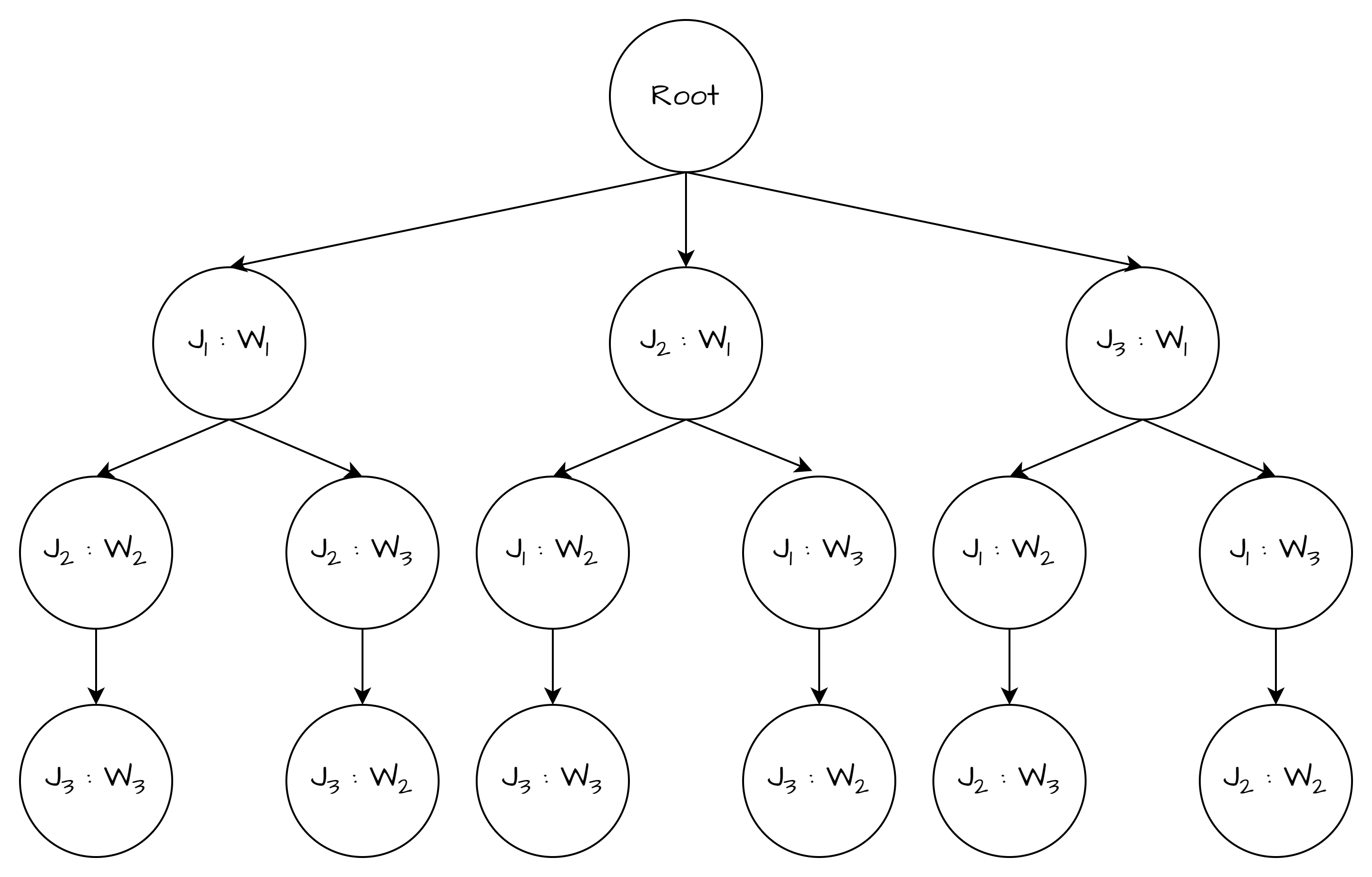}}
\caption{A State Space tree of the Illustrative Example peeked in the Section 5. We have considered a set of 3 Jobs, $\left\{\mathcal{J}_1,\mathcal{J}_2,\mathcal{J}_3\right\}$, and a set of 3 workers, $\left\{\mathcal{W}_1,\mathcal{W}_2,\mathcal{W}_3\right\}$.}
\end{figure*}
Here, as mentioned in Section 4, we can implicate any of the suitable Branch and Bound Technique [17]. Though, in each of the cases, the Total Cost turns out to be 15. Therefore, considering all the possible solutions,
\begin{equation*}
argmin{\left(\sum_{i=0}^{2}\left(\sum_{j=0}^{2}\left({\hat{\mathfrak{C}}}_{\left(i,\ \ \ j\right)}\right)\right)\right)}=15
\end{equation*}
\section{Conclusion}
Figure 3 shows a comparative plot on the solution of a Balanced Assignment Problem, on the metrics of Computational Complexity on the Dimensions of Time. It is frequently estimated using the number of primary level functions conducted by the algorithm, with the assumption that each basic procedure ends up taking a fixed duration of time to complete. Because the runtime [18] of a methodology might very well differ for various components that are the same magnitude, the very worst time complexity, which refers to the highest amount of time required for input data of a suitable magnitude, is commonly considered. 
\begin{figure}[t]
\centerline{\includegraphics[width = \linewidth]{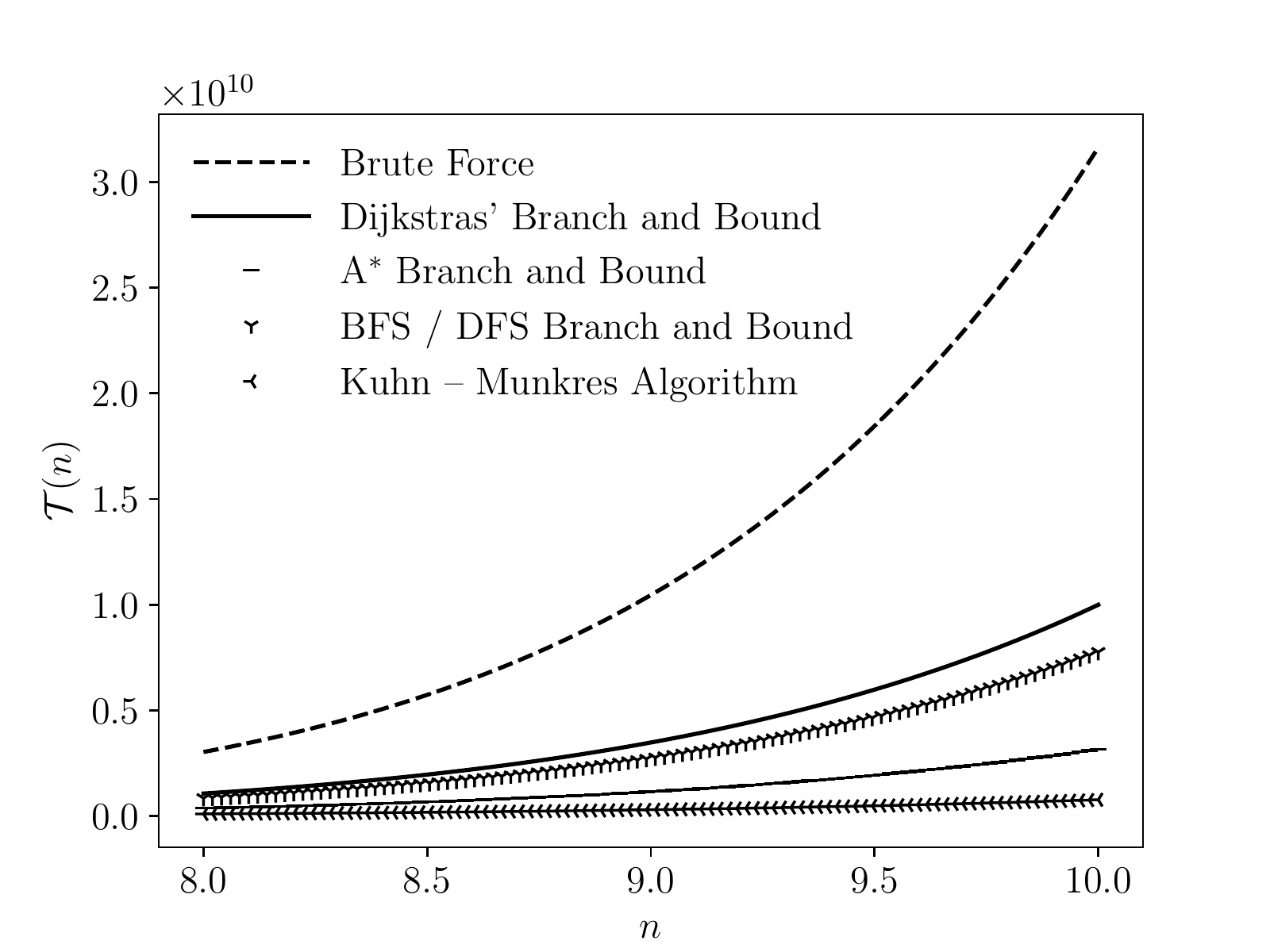}}
\label{fig1}
\caption{Comparative Plot}
\end{figure}
To Sum up, the most efficient solution is being catered by the A$^*$ Branch and Bound. Some, future scope of research includes, 
\begin{enumerate}
\item Improvising the Branch and Bound Method, by some more efficient Search. 
\item Improvising the Hungarian Algorithm. 
\end{enumerate}

\end{document}